\documentclass[twocolumn]{aastex631}
\usepackage{showyourwork}

\usepackage{xspace}

\usepackage{xcolor, fontawesome}
\definecolor{twitterblue}{RGB}{64,153,255}

\newcommand{\twitter}[1]{\href{https://twitter.com/#1}{\textcolor{twitterblue}{\faTwitter}\,\tt \textcolor{twitterblue}{@#1}}}

\newcommand{\github}[1]{\href{https://github.com/#1}{\textcolor{black}{\faGithub}\,\tt \textcolor{black}{#1}}}

\newcommand{\githubicon}{{\color{black}\faGithub}}

\newcommand{\tess}{\textit{TESS}}
\newcommand{\sname}{V1298~Tau\xspace}
\newcommand{\allplanets}{V1298~Tau~bcde\xspace}
\newcommand{\planetb}{V1298~Tau~b\xspace}
\newcommand{\planetc}{V1298~Tau~c\xspace}
\newcommand{\planetd}{V1298~Tau~d\xspace}
\newcommand{\planete}{V1298~Tau~e\xspace}
\newcommand{\planetknown}{V1298~Tau~bcd\xspace}
\newcommand{\rearth}{$R_\oplus$\xspace}
\newcommand{\exoplanet}{\texttt{exoplanet}\xspace}

\submitjournal{ApJL}

\shorttitle{V1298 Tau with \tess}
\shortauthors{Feinstein et al.}

\begin{document}

\title{V1298~Tau with TESS: Updated Ephemerides, Radii, and Period Constraints from a Second Transit of V1298~Tau~e}

\author[0000-0002-9464-8101]{Adina~D.~Feinstein}
\altaffiliation{NSF Graduate Research Fellow}
\affiliation{Department of Astronomy and Astrophysics, University of Chicago, Chicago, IL 60637, USA}

\author[0000-0001-6534-6246]{Trevor J.\ David}
\affiliation{Center for Computational Astrophysics, Flatiron Institute, New York, NY 10010, USA}
\affiliation{Department of Astrophysics, American Museum of Natural History, New York, NY 10024, USA}

\author[0000-0001-7516-8308]{Benjamin~T.~Montet}
\affiliation{School of Physics, University of New South Wales, Sydney, NSW 2052, Australia}
\affiliation{UNSW Data Science Hub, University of New South Wales, Sydney, NSW 2052, Australia}

\author[0000-0002-9328-5652]{Daniel Foreman-Mackey}
\affiliation{Center for Computational Astrophysics, Flatiron Institute, New York, NY 10010, USA}

\author[0000-0002-4881-3620]{John~H.~Livingston}
\affiliation{Department of Astronomy, University of Tokyo, 7-3-1 Hongo, Bunkyo-ku, Tokyo 113-0033, Japan}

\author[0000-0003-3654-1602]{Andrew~W.~Mann}
\affiliation{1Department of Physics and Astronomy, The University of North Carolina at Chapel Hill, Chapel Hill, NC 27599, USA}

\correspondingauthor{Adina~D.~Feinstein;\\ \twitter{afeinstein20}; \github{afeinstein20};} \email{afeinstein@uchicago.edu}

\begin{abstract}

\sname is a young (20--30~Myr) solar-mass K star hosting four transiting exoplanets with sizes between $0.5 - 0.9 R_J$. Given the system's youth,  it provides a unique opportunity to understand the evolution of planetary radii at different separations in the same stellar environment. \sname was originally observed 6 years ago during \textit{K2} Campaign 4. Now, \sname has been re-observed during the extended mission of NASA's Transiting Exoplanet Survey Satellite (\tess). Here, we present new photometric observations of \sname from the 10-minute \tess\ Full-Frame Images. We use the \tess\ data to update the ephemerides for \allplanets as well as compare newly observed radii to those measured from the \textit{K2} light curve, finding shallower transits for \planetknown in the redder \tess\ bandpass at the $1-2\sigma$ level. We suspect the difference in radii is due to starspot-crossing events or contamination from nearby faint stars on the same pixels as \sname. Additionally, we catch a second transit of \planete and present a new method for deriving the marginalized posterior probability of a planet's period from two transits observed years apart. We find the highest probability period for \planete to be in a near 2:1 mean motion resonance with \planetb which, if confirmed, could make \allplanets a 4 planet resonant chain. \sname is the target of several ongoing and future observations. These updated ephemerides will be crucial for scheduling future transit observations and interpreting future Doppler tomographic or transmission spectroscopy signals.

\end{abstract}




\section{Introduction} \label{sec:intro}
Planetary radii are expected to evolve over time, due to a variety of endogenous and exogenous physical processes, such as gravitational contraction, atmospheric heating and mass-loss, and core-envelope interactions \citep[e.g.][]{OwenWu2013, Lopez2013, Jin2014, ChenRogers2016, Ginzburg2018}. The most dramatic changes are believed to occur at early stages, when planets are still contracting and radiating away the energy from their formation, and when host stars are heating planetary atmospheres with high levels of X-ray and ultraviolet radiation. Since the size evolution of any individual planet is believed to be slow relative to typical observational baselines, the best way to make inferences about the size evolution of exoplanets is by measuring the sizes of large numbers of planets across a range of ages. 

NASA's Transiting Exoplanet Survey Satellite \citep[\tess;][]{Ricker2015} has made significant inroads toward this objective. \tess's observations of $\sim 90 \%$ of the sky have allowed for exoplanet transit searches around stars ranging from the pre-main sequence to the giant branch. It is through targeted surveys of young stars such as the THYME \citep[e.g.][]{Newton2019}, PATHOS \citep[e.g.][]{Nardiello2020}, and CDIPS \citep[e.g.][]{Bouma2020} surveys, along with case studies of individual systems \citep[e.g.][]{benatti19, Plavchan2020, Hedges2021, Zhou2021} that the timeline for planetary radii evolution can be pieced together. 

The \sname planetary system is one particularly valuable benchmark for understanding the size evolution of exoplanets. \sname is a pre-main sequence, approximately solar-mass star that was observed in 2015 by NASA's \textit{K2} mission \citep{Howell2014}. Analysis of the \textit{K2} data revealed the presence of four transiting planets, all with sizes between that of Neptune and Jupiter \citep{David2019a, David2019b}. There are no other known examples of exoplanetary systems with so many planets larger than Neptune interior to 0.5~au, despite the high completeness of the \textit{Kepler} survey to large ($>5$~\rearth), close-in planets. This observation raises the possibility of a causal connection between the extreme youth of \sname and the uncommonly large sizes of its planets.  

The youth of \sname was initially established on the basis of its strong X-ray emission \citep{Wichmann1996}, high photospheric lithium abundance \citep{Wichmann2000}, and proper motion measurements \citep{frink1997}. An additional recent blind search for co-moving stars using Gaia DR1 astrometry data found \sname was co-moving with 8 other stars \citep[Group 29 in][]{Oh2017}. \cite{Luhman2018} conducted a kinematic study of the Taurus star-forming region using Gaia DR2 and found new members of this group. With this new sample, they derived an age of $\sim$~40~Myr. However, more recent analyses based on Gaia EDR3 astrometry suggests \sname may belong to either the D2 or D3 subgroups of Taurus, both of which have estimated ages $\lesssim$10~Myr \citep{gaidos21, Krolikowski2021}. Other studies focused specifically on the \sname system have estimated its age to be 23$\pm$4~Myr from comparison with empirical and theoretical isochrones \citep{David2019b}, or 28$\pm$4~Myr from isochrone fitting to the \citet{Luhman2018} Group 29 membership list given Gaia EDR3 data \citep{johnson21}. While the precise age of \sname remains uncertain, most estimates fall in the 10--40~Myr range and we adopt $t \approx$~20--30~Myr.

Given the system's youth and potential to reveal information about the initial conditions of close-in planetary systems \citep[e.g.][]{Owen2020,Poppenhaeger2021}, \sname has been the target for extensive follow-up observations. These include efforts to constrain planet masses with radial velocities \citep{Beichman2019,suarez21}, measure the spin-orbit alignments of planet c \citep{Feinstein21} and planet b \citep{johnson21, gaidos21}, measure or constrain atmospheric mass-loss rates for the innermost planets \citep{Schlawin21, Vissapragada21}, and an approved program to study the planetary atmospheres using the James Webb Space Telescope \citep[JWST;][]{Desert2021}.

Here we report on newly acquired \tess\ observations of \sname which help to refine the orbital ephemerides of the transiting planets and enable comparison of the planet sizes inferred from two different telescopes with different bandpasses (\tess\ and \textit{Kepler}). The goal of this letter is to provide a quick analysis of the new \tess\ data to help improve the transit timings for follow-up observations being performed by the community. We describe the observations and light curve extraction in Section~\ref{sec:observations}. In Section~\ref{sec:analysis}, we present our light curve modeling and method for computing the marginalized posterior probability of a transiting planet's period from two transits observed with a large time gap. In Section~\ref{sec:radii}, we discuss the differences in measured transit parameters between \textit{K2} and \tess\ data. We conclude in Section~\ref{sec:conclusions}.

\begin{figure}[t!]
\begin{center}
\includegraphics[width=0.46\textwidth,trim={0.25cm 0 0 0}]{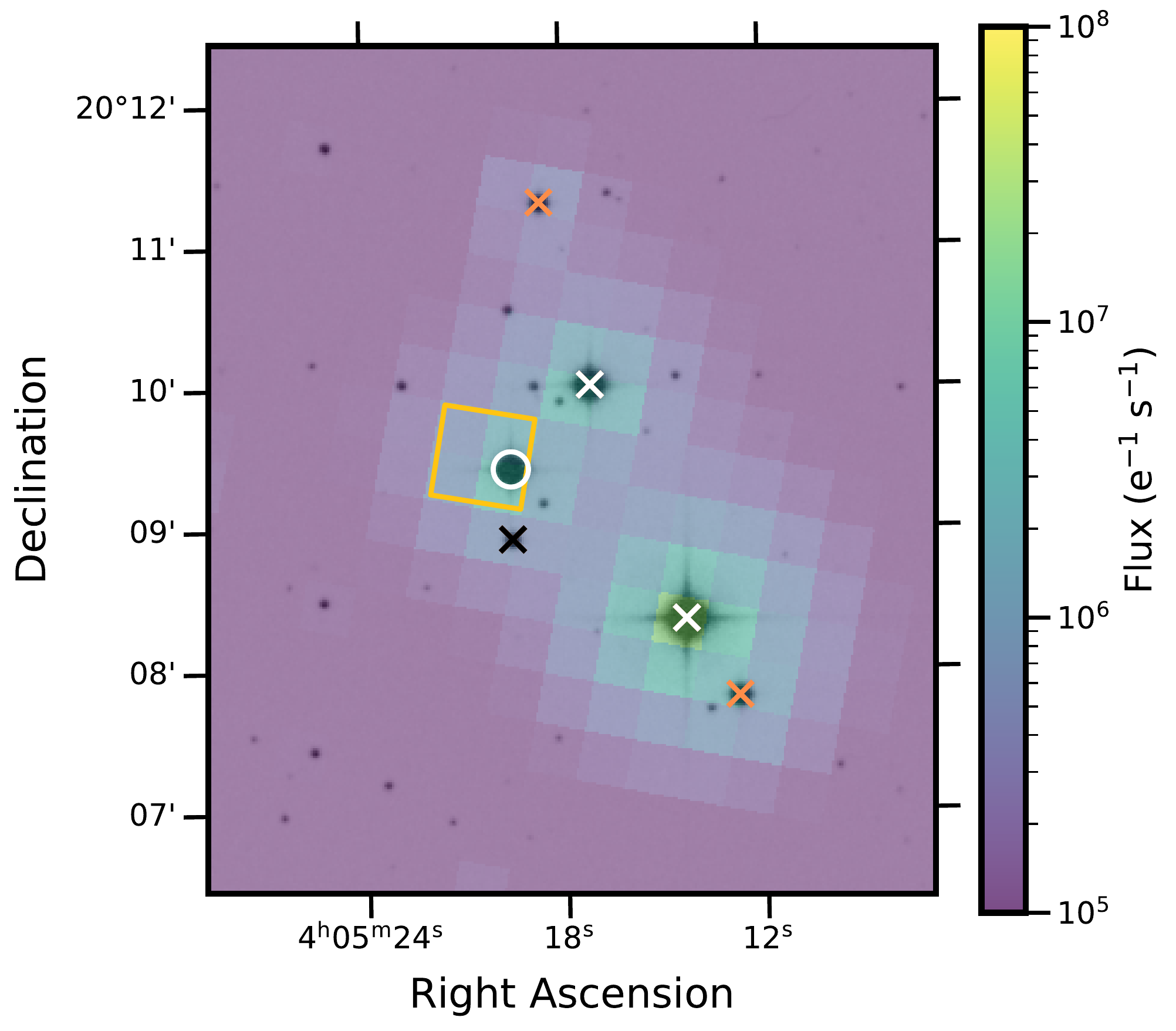}
\caption{The \tess\ \texttt{tica} FFI target pixel file (TPF) overlaid with a sky image of \sname taken with the Digitized Sky Survey (DSS) r-band. \sname is highlighted by the white circle; nearby sources with \tess\ magnitudes $< 14$ are marked with x's. The two stars with white x's (TICs 15756226 and 15756240) were simultaneously fit during our PSF-modeling. The one star with black x (TIC 15756236) is a bright nearby source that was not included in our PSF-modeling. While aperture photometry would be feasible for this system (yellow square), we found fitting three point-spread functions to the brightest stars extracted the cleanest light curve for \sname. \href{https://github.com/afeinstein20/v1298tau\_tess/blob/main/src/figures/tpf.py}{\githubicon}} \label{fig:tpf}
\end{center}
\end{figure}

\section{TESS Observations} \label{sec:observations}

During its Extended Mission Cycle 4, \tess\ is re-observing many of the previous \textit{K2} fields. \sname (TIC 15756231) was observed by \tess\ in Sectors 43 (UT 16 Sep 2021 to UT 12 Oct 2021) and 44 (UT 12 Oct 2021 to UT 06 Nov 2021). For Sectors 43 and 44, we used the 2-minute light curve created by the Science Processing Operations Center pipeline \citep[SPOC;][]{jenkinsSPOC2016} and binned the data down to 10-minute cadence.

\begin{figure*}[hbtp]
\begin{center}
\includegraphics[width=0.9\textwidth,trim={0.25cm 0 0 0}]{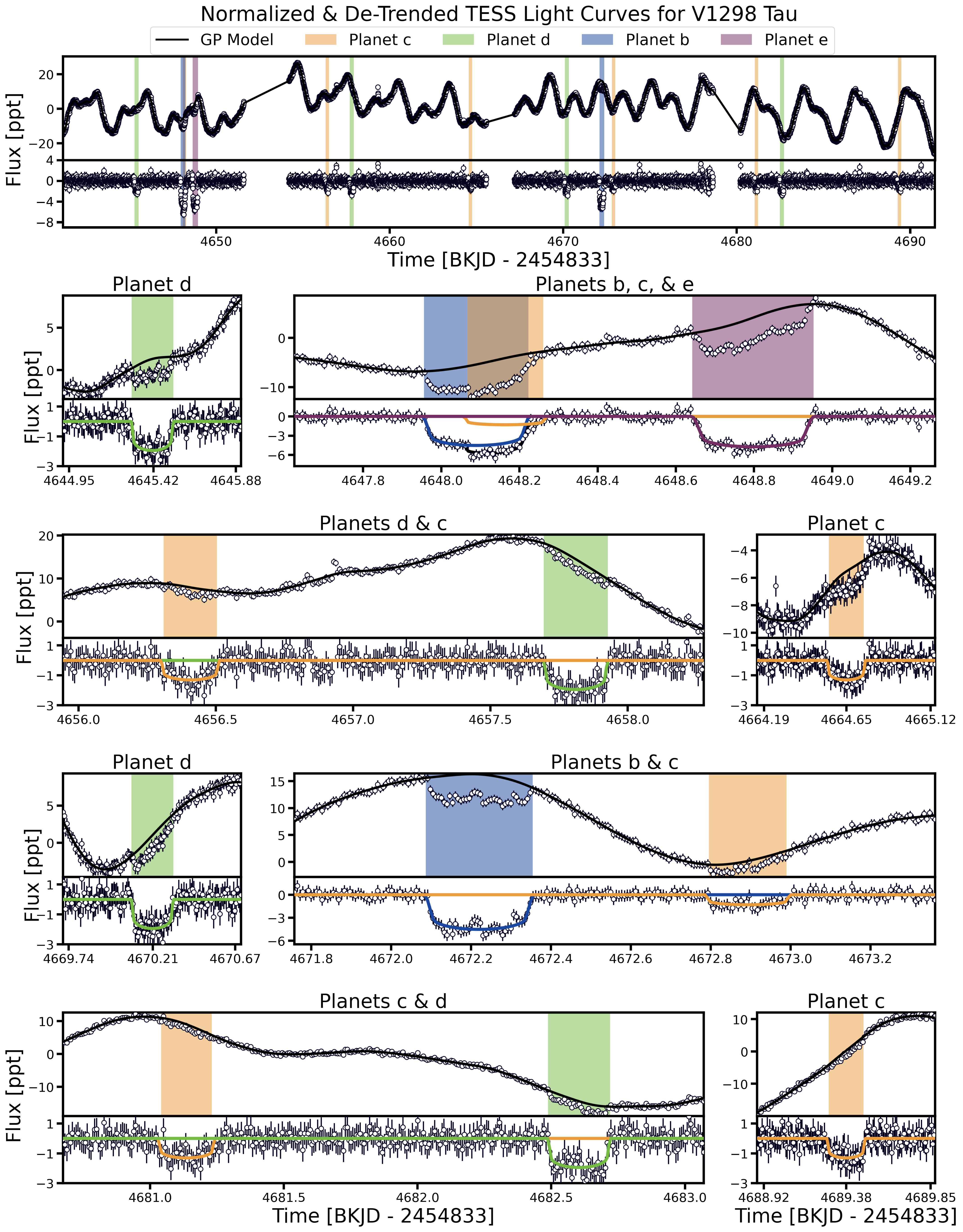}
\caption{\sname extracted light curve from the SPOC-processed light curve for Sectors 43 and 44, with transits of \allplanets highlighted by color. Each subplot contains the raw, normalized \tess\ flux (top) and the GP model removed flux (de-trended flux; bottom). Top row: extracted light curve with over plotted with our best-fit GP model for stellar variability (black). Bottom three rows: zoomed-in regions around the transits present in the \tess\ data. The GP best-fit model is over-plotted on the raw, normalized flux. The best-fit transit models are over-plotted on the de-trended flux. For overlapping transits (sub-panel ``Planets b, c, \& e''), the sum of the transits is blotted in black. \href{https://github.com/afeinstein20/v1298tau\_tess/blob/main/src/figures/transits.py}{\githubicon}} \label{fig:transits}
\end{center}
\end{figure*}

We compared these new light curves to our original FFI light curves. We created our initial light curves from the \texttt{tica}-processed FFIs by modeling the point-spread function (PSF) of \sname and the two nearby bright sources (white x's in Figure~\ref{fig:tpf}), following the PSF modeling routine in \cite{feinstein19}.\footnote{Our PSF-modeled light curves are available \href{https://github.com/afeinstein20/v1298tau_tess/tree/main/lightcurves}{here}.} In summary, we calculated and maximized the likelihood value of seven parameters per each Gaussian: the $x$ and $y$ width, 2D position, amplitude, a rotational term, and a background term. The Gaussian fits are allowed to vary at each time step. Aperture photometry (example square aperture shown in Figure~\ref{fig:tpf}) provided a light curve with more systematics and scatter. We found that modeling the three brightest stars simultaneously, including \sname, with a 2D Gaussian created the least contaminated light curve. Our extracted light curve is shown in the top row of Figure~\ref{fig:transits}.

\section{Analysis} \label{sec:analysis}

We simultaneously modeled the transits of \allplanets and the stellar variability using the open-source packages \exoplanet \citep{exoplanet2019, exoplanet2021} and \texttt{PyMC3} \citep{Salvatier16}. Transit timings were originally identified using updated ephemerides from \textit{Spitzer} (Livingston et al. in prep) and we account for potential transit timing variations (TTVs) in our model using \texttt{exoplanet}'s \texttt{TTVOrbit} function. This functionality allows us to fit the individual transit time per each transit, while all other orbital and planetary parameters remain the same.

\texttt{exoplanet.TTVOrbit} worked well when fitting \planetknown, due to there being multiple transits. However, with \planete being a single-transit event, we found the GP model with TTVs optimized our hyperparameters to accommodate for transits where there were none; as such, we used a model without accounting for TTVs to fit the parameters for \planete. All other transit parameters (presented in Table~\ref{tab:table}) were initialized using values from \cite{David2019a}. We assumed a quadratic limb darkening law, following the reparameterization described by \cite{kipping13}; this method allows for an efficient and uniform sampling of limb-darkened transit models.


Since the \texttt{tica} FFIs do not provide an error estimate, we fit for flux errors within our Gaussian Process (GP) model. We define the flux error as

\begin{equation}
    \sigma_y = e^{ln(\sigma_l)} + y^2 e^{2 ln(\sigma_j)}
\end{equation}

where $y$ is the normalized flux array about zero, and $\sigma_l$  and $\sigma_j$ are used to define the light curve noise and in-transit jitter, which is designed to capture the added noise produced by starspot crossing events. $\sigma_l$  and $\sigma_j$ are also used as the first and second terms in our rotation model, which we defined as a stochastically-driven, damped harmonic oscillator, defined by the \texttt{SHOTerm} in \texttt{celerite2} \citep{dfm17}.

We modeled the background within our GP. We defined a quadratic trend with respect to time for varying the background flux, where each polynomial coefficient was drawn from a normal distribution. Then, we generated a Vandermonde matrix ($A$) of time. This is a way of introducing a polynomial least-squares regression with respect to time. The final background flux was calculated by taking $bkg = A \cdot trend$.  

We performed an MCMC sampling fit to each parameter.\footnote{A Jupyter notebook detailing our model can be found \href{https://github.com/afeinstein20/v1298tau\_tess/blob/main/notebooks/TESS\_V1298Tau.ipynb}{here}.} We ran 3 chains with 500 tuning steps and 5000 draws. We discarded the tuning samples from the posterior chains before calculating our best-fit parameters. Our results are presented in Table~\ref{tab:table}, along with our selected priors for each parameter we fit. These results are consistent with our original \texttt{tica}-processed point-spread function modeled light curves. We verified our chains converged via visual inspection and following the diagnostic provided by \cite{Geweke92}.

We present our final GP model for stellar variability, planet transits, and best-fit transit models in Figure~\ref{fig:transits}. There is a $\sim 1\%$ flare at \tess\ BKJD $\approx$ 4659.18 that we do not fit.

\subsection{Constraining \planete's Period}

\sname (EPIC~210818897) was observed during Campaign 4 of the \textit{K2} mission. There was a single transit of \planete in the original \textit{K2} data, which occurred roughly in the middle of the campaign. Since no other transits were detected, this provides a lower period limit of 36~days. Additionally, there was only 1 transit of \planete between the two \tess\ sectors, which provides a new lower limit of 42.7~days. Using the original transit timing from \textit{K2} and this new transit timing from \tess, we developed a new method for constraining the period of \planete. For this analysis, we used the \texttt{EVEREST 2.0} \citep{luger18} version of the \textit{K2} light curve. 

Determining the period of a planet from two transits with a significant time gap between surveys has previously been constrained by fitting for orbital periods using MCMC sampling, phase-folding all available transits on the derived transit times and periods, and computing a reduced-$\chi^2$ fit to a flat line \citep{becker19}. Orbital periods providing a match to a flat line with a likelihood exceeding some threshold are then ruled out. In our new method, we fit transit models of many discrete periods at each step of the MCMC sampler, rather than post-processing from our posterior.

First, we de-trended a localized 1-day region around the transit midpoint of \planete in the \textit{K2} and \tess\ light curves, assuming a constant transit depth and allowing the other transit parameters, $\theta$ to vary. Then, we fit a discrete period model, allowing all other transit parameters to vary. We set the GP model to sample over discrete periods ranging from $38 - 56$~days. We fit for $\theta$ assuming a constant transit depth between the \textit{K2} and \tess\ observations. We assumed there is no correlation between the other transit parameters and the period we are fitting for. For each step in our MCMC fit, we compute all possible periods

\begin{equation}\label{eq:period}
    P = \frac{1}{q} \left(T_{mid,TESS} - T_{mid, K2}\right)
\end{equation}

where $T_{mid}$ are the transit midpoints from \textit{K2} and \tess\ and $q$ is an integer representing a specific harmonic. We assume a uniform prior, i.e. we have no prior preference for a specific harmonic. At each step of the sampling process, we compute a new light curve with different orbital periods, given by Equation~\ref{eq:period}. The log likelihood of the new light curves models are calculated as

\begin{equation}
    \textrm{log} \mathcal{L}_q = \left[ \textrm{log}\, p \left( X | \theta^k, q^k = n \right) \right]_n
\end{equation}

where $X$ is the \tess\ light curve and $n$ is the period being tested. We additionally calculate the sum of all log likelihoods for each period 

\begin{equation}
    \textrm{log} \mathcal{L} = \textrm{log}\, \Sigma_q\, p(q)\, p(X|\theta^k, q) 
\end{equation}

The summation of all log likelihoods is used to compute the posterior likelihood for each sampled value of $q$. This analysis assumes a circular orbit for planet e and uses stellar density constraints via priors on the stellar mass and radius.\footnote{A Jupyter notebook detailing our model for constraining the period for \planete can be found \href{https://github.com/afeinstein20/v1298tau\_tess/blob/main/notebooks/V1298Tau\_e.ipynb}{here}.}

\begin{figure}[h]
\begin{center}
\includegraphics[width=0.44\textwidth,trim={0.25cm 0 0cm 0}]{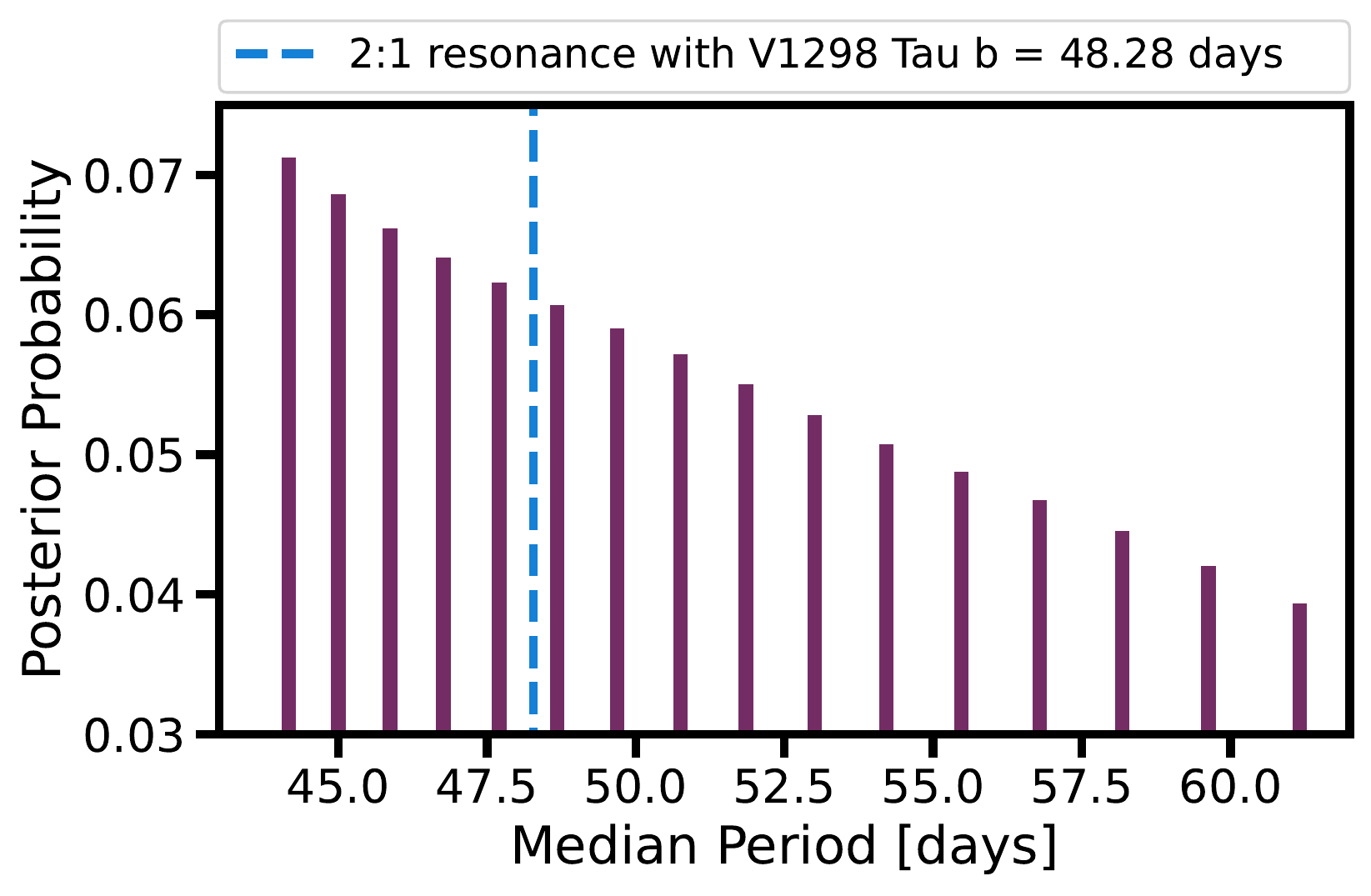}
\caption{Our calculated posterior probability to constrain the period of \planete using transit timings from \textit{K2} and \tess. We tested discrete periods from $q=38-56$\,days and find the most likely period to be 44.17~days. The 2:1 resonance (48.28~days) with \planetb is plotted as the dashed vertical line. \href{https://github.com/afeinstein20/v1298tau\_tess/blob/main/src/figures/period\_e.py}{\githubicon}} \label{fig:period_e}
\end{center}
\end{figure}

We ran 3 chains with 500 tuning steps and 5000 draws. We discarded the tuning steps before our analysis. Our results are presented in Figure~\ref{fig:period_e}, where we plot the median period for each tested harmonic against the posterior probability of each harmonic.

We find the most likely period of \planete to be 44.17~days. We provide all tested periods and posterior probabilities in Table~\ref{tab:e}. Our period estimate is at a 4-$\sigma$ disagreement with the period measured in a potential radial velocity signal for \planete presented in \cite{suarez21}. This derived period estimate suggests that \planete is in a near 2:1 mean motion resonance with \planetb. If the period of \planete is confirmed to be within the presented range, this could indicate that \allplanets are in a nearly 4-planet resonant chain.

Independent ground-based monitoring of the system may be able to observe another transit of the outermost known planet in this system. Using the Transit Service Query Form on the NASA Exoplanet Archive \citep{Akeson2013}, we provide several potential transit midpoint events for all fitted periods in Table~\ref{tab:e}.

\section{Differences in Measured Radii} \label{sec:radii}

We compare the differences in transit $R_p/R_\star$ between the \textit{K2} and \tess\ data in Figure~\ref{fig:compare}. We masked regions in the light curve where transits overlapped. The residuals of the \tess\ light curve with our model (color) are plotted as well. For \planetknown, the transit radii are smaller in the \tess\ data, while only the measured radius for \planete is larger (Figure~\ref{fig:compare}, bottom panel).

The error bars from our MCMC fit on the radii of the planets are smaller than that provided by \cite{David2019b}. We initialized our MCMC to fit the transit depths with a Gaussian distribution around the fitted values from \cite{David2019b} with a standard deviation of 0.1 (Table~\ref{tab:table}). The smaller errors could be due to the higher cadence of the \tess\ data (10-minutes vs. 30-minutes) or due to larger spot-crossing events in \textit{K2}. Larger spot-crossings would result in a greater uncertainty of the transit depth, and this is potentially evident in comparing the transit depth and shape for \planete (Figure~\ref{fig:compare}).

\subsection{Radii of \planetknown}

\begin{figure*}[!htb]
\begin{center}
\includegraphics[width=\textwidth,trim={0.25cm 0 0 0}]{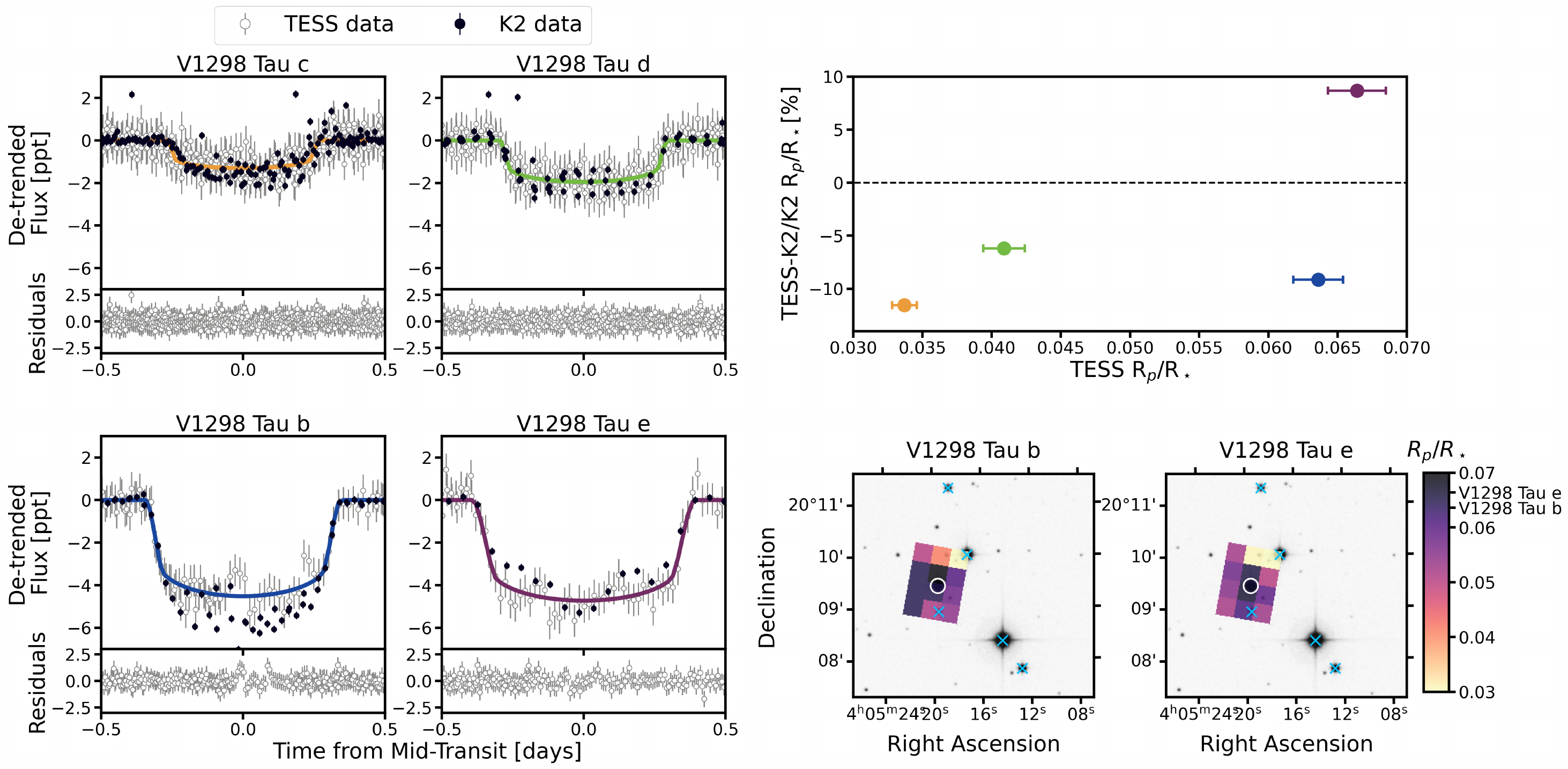}
\caption{Left: Phase-folded \tess\ data (gray) with the new best-fit model (color) compared to the original \textit{K2} data (black). The residuals between the \tess\ data and each fit are shown underneath. Top right: The percent difference in measured $R_p/R_\star$ from \textit{K2} vs. \tess\ $R_p/R_\star$. A dashed line is shown at 0\% to help visually differentiate between measured increases and decreases in planetary radii. The transit depths for \planetknown are shallower in the new \tess\ data, while the transit depth for \planete is deeper. Bottom right: The change in $R_p/R_\star$ as a function of pixel for \planetb (left) and \planete (right) overlaid with a sky image of \sname taken with the Digitized Sky Survey (DSS) r-band. We speculate the variation in transit depths could be due to contamination from nearby bright stars or starspot crossing events. \href{https://github.com/afeinstein20/v1298tau_tess/blob/main/src/figures/dilution_check.py}{\githubicon}} \label{fig:compare}
\end{center}
\end{figure*}

A shallower transit depth at redder wavelengths is supported by the dusty outflow model presented in \cite{wang19}, while transits in the optical probe lower atmospheric pressures, resulting in larger transit depths \citep{gao20}. Therefore, the difference in transit depths could be due to the difference in bandpass wavelength coverage between \textit{Kepler} \citep[400-900~nm;][]{Howell2014} and \tess\ \citep[600-1000~nm;][]{Ricker2015}. 

The ability of a planet to retain hazes/transition hazes is negatively correlated with its equilibrium temperature, $T_{eq}$, internal temperature $T_{int}$, and positively correlated with mass, $M_{core}$. \allplanets have calculated $T_{eq} < 1000$\,K, assuming an albedo~=~0 \citep{David2019a}. Young planets are believed to have high $T_{int}$ due to ongoing gravitational contraction \citep{gu04}. The combination of these two parameters make these planets more likely to host extended high-altitude hazes in their atmospheres, while outflow winds lead to the formation of transition hazes \citep{gao20}. 

However, it is more likely we are seeing either contamination/dilution from another nearby star or the presence of starspots. As highlighted by a black x in Figure~\ref{fig:tpf}, there is a bright (\tess\ magnitude $<$ 14) just next to \sname. Since we did not include this source in our light curve PSF-modeling, it is possible there is some light from this source in our data, therefore making the transits of \planetknown appear $\sim 10$\% shallower in our \tess\ data than the original \textit{K2} measurements. The \tess\ Input Catalog \citep{stassun18} lists a contamination value of 0.315 for \sname, which could sufficiently produce the decrease in transit depths presented here. 

We check for signs of dilution by creating light curves for individual pixels around \sname and measure the transit depths of \planetb and \planete (Figure~\ref{fig:compare}). We localized a 1-day window around each transit and computed the a $\chi^2$-fit using a transit model computed with \texttt{batman} \citep{Kreidberg15} and an underlying 2\textsuperscript{nd} order polynomial. We find the transit depths to decrease falling off of the pixels centered on \sname. While fitting the \tess\ PSF with a 2D Gaussian function is a reasonable approximation, it is not the perfect model. It is possible that this light curve is diluted from nearby stars, including TIC~15756226, which is the closest (separation $= 49.15\arcsec$) star with $T_\textrm{mag} = 13.09$.

A change in transit depth could additionally be due to starspots, either from a nearby source or from the surface of \sname. In the context of starspots on \sname, both starspot/active region crossing events, where the planets directly transit over these inhomogeneities, or asymmetric starspots/active region distributions located off the transit chord could lead to differences in transit depths and shapes. In the case of starspot crossing events, we would expect to see added variability to the transit shape. Assuming we are not underestimating our error bars, this is readily seen in the transit of \planetb, both in the \tess\ and \textit{K2} data (Figure~\ref{fig:compare}).

\subsection{Radius of \planete}

Contrary to \planetknown, we measure a transit radius that is $\sim 3\sigma$ larger in the new \tess\ data than what was found in the original \textit{K2} data \citep{David2019a}. The difference in radii could be consistent with a large scale height, low mean-molecular weight atmosphere around \planete \citep{deMooij12}. It is also possible the atmosphere of \planete is dominated by species with stronger absorption features at longer wavelengths, such as CO, H$_2$O, and CH$_4$ \citep{carter09}. 

We find it is more likely that the original single transit of \planete was filled-in via spot-crossing events, making it appear shallower in the \textit{K2} observations. Young stars are known to have very high spot coverage, anywhere from 30-80\% \citep[][]{grankin99, gully17, feinstein20}. It is therefore likely the surface of \sname is dominated by stellar inhomogeneities. This hypothesis is further strengthened by comparing the transit shape between the two data sets (Figure~\ref{fig:compare}). The center of the \textit{K2} transit is deeper than the edges and is consistent with the most recently observed transit depth. The lower contrast in starspot signals at longer wavelengths could potentially explain why the transit depth is consistent deeper in the \tess\ observations. Additionally, there is noise in-transit in the \tess\ data that could potentially be more starspot crossing events.

Future \tess\ 20-second and 2-minute data may have the temporal resolution needed to yield insight into if there is evidence of starspot crossing events. Our transit of \planete from the FFIs shows some structure. At the 10-minute cadence, it is hard to rule out noise as the source of this structure. However if there is such evidence of starspot crossings in the higher-cadence \tess\ observations, it would be interesting if any of the events dilutes the transit enough to result in a similar transit depth to that which was seen in \textit{K2}. 

\section{Conclusions} \label{sec:conclusions}

We present updated ephemerides for all four known planets in the \sname system. Our GP model accounts for TTVs for \planetknown. The transit timings for \planetc deviates from a linear ephemeris by -30 to 30 minutes, and \planetd deviates by -5 to 5 minutes. Additionally, we note the transits of \planetknown occur 1.92 later, and 5.83 and 4.72 hours earlier than what is expected if we extrapolate forward the ephemerides from \cite{David2019b}.

We detected a second transit of \planete; this new transit time in combination from the transit observed with \textit{K2} allowed us to place tighter constraints on the period of the outermost planet. Our revised radius for planet e makes it now the largest planet in the system and extends an intriguing size--separation correlation in \sname such that planet size increases monotonically with separation.

We find the transit depths of \planetknown as observed by \tess\ are shallower than those observed by \textit{K2} by $1-2\sigma$, with the exception of \planete, which is $\sim 3\sigma$ larger. While this could possibly be due to ongoing dusty outflows that make the transit depth appear shallower, it is more likely the differences are due to starspot crossing events, asymmetric starspots off the transit chord, or contamination from nearby faint stars in the same \tess\ pixels are \sname. Modeling potential starspot crossing events could be accomplished using the 2-minute and 20-second cadence light curves, which will be available in the coming months.

The youth of these planets could additionally favor hosting haze-dominated atmospheres. However, without mass estimates for \allplanets, it is difficult to determine if this is the cause of the different transit depths measured between the \textit{K2} data and our presented work. Radial velocity mass measurements are challenging for young planets due to underlying stellar activity.

Through an intensive radial velocity campaign, \cite{suarez21} presented a new mass detection for \planetb and \planete. Our updated radius estimate for \planetb in combination with the mass estimate provided by \cite{suarez21} would yield a density of $1.29$~g/cm\textsuperscript{3}, which is slightly higher than what was originally reported. Our updated radius for \planete would yield a lower density of $2.04$~g/cm\textsuperscript{3}, which is still within a 1-$\sigma$ agreement with \cite{suarez21}. However, the period estimate for \planete is at a 4-$\sigma$ disagreement for the period estimate provided in this study and at a 2-$\sigma$ disagreement with the minimum period constraint provided by this new \tess\ data.

A more promising approach to measuring the masses would be through TTVs \citep{agol18}. A full analysis of system parameters and TTVs from both the \textit{K2} and \tess\ light curves, and additional \textit{Spitzer} transit photometry will be presented in Livingston et al. (in prep). Additional transits at longer wavelengths or simultaneous multi-band photometry or spectroscopy could corroborate the potential of constraining the properties of these young atmospheres.

\vspace{0.5mm}

We thank Rodrigo Luger for developing \texttt{showyourwork!} \citep{luger21} and helping us debug this letter. We thank Chas Beichman, Sarah Blunt, Jacob Bean, and Darryl Seligman for helpful comments on our \tess\ proposal (DDT 036) and thoughtful conversations. We thank our anonymous referee for their thoughtful insights which improved the quality of this manuscript. ADF acknowledges support from the National Science Foundation Graduate Research Fellowship Program under Grant No. (DGE-1746045).

This research has made use of NASA's Astrophysics Data System Bibliographic Services. This research made use of Lightkurve, a Python package for \textit{Kepler} and \tess\ data analysis \citep{lightkurve}.


\begin{deluxetable*}{l r r r r}[hbtp]
\tabletypesize{\footnotesize}
\tablecaption{\sname light curve fitting results and predicted ground-based transit midpoint events for \planete. \label{tab:table}}
\tablehead{\\
\hline\
\textit{Star} & \textit{Value} & \textit{Prior} &  & \\
\hline
$R_\star [R_\odot]$ & $1.33_{-0.03}^{+0.04}$ & $\mathcal{G}(1.305, 0.07)$ & & \\
$M_\star [M_\odot]$ & $1.095_{-0.047}^{+0.049}$ & $\mathcal{G}(1.10, 0.05)$& & \\
$u_1$ & $0.32_{-0.19}^{+0.20}$ & $\mathcal{U}[0, 1]$ in $q_1$ & & \\
$u_2$ & $0.16_{-0.29}^{+0.31}$ & $\mathcal{U}[0, 1]$ in $q_2$ & & \\
$P_{rot}$ [days] & $2.97_{-0.04}^{+0.03}$ & $\mathcal{G}(\textrm{ln} 2.87, 2)$ & & \\
ln($Q_0$) & $0.72_{-0.21}^{+0.24}$ & $\mathcal{H}(\sigma=2)$ & & \\
$\Delta Q_0$ & $4.09 \pm 1.01$ & $\mathcal{G}(0, 2)$ & & \\
f [ppt] & $0.85_{-0.19}^{+0.11}$ & $\mathcal{U}[0.1, 1]$ & & \\
\hline\
\textit{Light Curve} & \textit{Value} & \textit{Prior} &  & \\
\hline
$\mu$ & $-1.66_{-9.28}^{+9.42}$& $\mathcal{G}(0, 10)$ & &\\
ln($\sigma_l$) & $-2.499_{-4.463}^{+4.444}$ & $\mathcal{G}(ln(0.1\sigma_\textrm{flux}), 10)$ & & \\
ln($\sigma_j$) & $-1.36_{-5.06}^{+4.96}$ & $\mathcal{G}(ln(0.1\sigma_\textrm{flux}), 10)$ & & \\
\hline\
\textit{Planets} & \textit{c} & \textit{d} & \textit{b} & \textit{e}\\
\hline
$T_0$ [BKJD - 2454833] & $4648.16636_{-0.00339}^{+0.00269}$ & $4645.41494_{-0.00157}^{+0.00172}$ & $4648.09023_{-0.00132}^{+0.00129}$ & $4648.79668_{-0.00114}^{+0.00121}$ \\
$P$ [days]               & $8.2438_{-0.0020}^{+0.0024}$ & $12.3960_{-0.0020}^{+0.0019}$ & $24.1315_{-0.0034}^{+0.0033}$ & 44.1699 \\
$R_p/R_\star$          &  $0.0337 \pm 0.0009$ & $0.0409_{-0.0015}^{+0.0014}$ & $0.0636 \pm 0.0018$ & $0.0664_{-0.0021}^{+0.0025}$ \\
Impact parameter, $b$                    &  $0.14_{-0.10}^{+0.14}$ & $0.19_{-0.13}^{+0.12}$ & $0.45_{-0.04}^{+0.05}$ & $0.48_{-0.07}^{+0.06}$ \\
T$_{14}$ [hours]        &  $4.66_{-0.43}^{+0.49}$ & $5.59_{-0.53}^{+0.57}$ & $6.42_{-0.61}^{+0.66}$ & $7.44_{-0.71}^{+0.79}$  \\
$R_p [R_\oplus]$       & $5.05 \pm 0.14$ & $6.13 \pm 0.28$ & $9.53 \pm 0.32$ & $9.94 \pm 0.39$\\
$R_p [R_J]$       & $0.45 \pm 0.01$ & $0.55 \pm 0.03$ & $0.85 \pm 0.03$ & $0.89 \pm 0.04$\\
TTVs [minutes] & $-0.41 \pm 25.38$  & $-0.12 \pm 4.08$ & --- & --- \\
\hline\
\textit{Priors} & \textit{c} & \textit{d} & \textit{b} & \textit{e}\\
\hline
$T_0$ [BKJD - 2454833] & $\mathcal{G}(4648.53,0.1)$ & $\mathcal{G}(4645.4,0.1)$ & $\mathcal{G}(4648.1,0.1)$ & $\mathcal{G}(4648.8,0.1)$ \\
log(P [\textrm{days}]) & $\mathcal{G}(\textrm{ln} 8.25, 1)$ & $\mathcal{G}(\textrm{ln} 12.40, 1)$ & $\mathcal{G}(\textrm{ln} 24.14, 1)$ & $\mathcal{G}(\textrm{ln} 36.70, 1)$\\
log(depth [ppt]) & $\mathcal{G}(\textrm{ln} 1.45, 0.1)$ & $\mathcal{G}(\textrm{ln} 1.90, 0.1)$ & $\mathcal{G}(\textrm{ln} 4.90, 0.1)$ & $\mathcal{G}(\textrm{ln} 3.73, 0.1)$ \\
Impact parameter, $b$ & $\mathcal{U}[0, 1]$ & $\mathcal{U}[0, 1]$ & $\mathcal{U}[0, 1]$ & $\mathcal{U}[0, 1]$ \\
T$_{14}$ [days] & $\mathcal{G}(\textrm{ln} 0.19, 1)$ & $\mathcal{G}(\textrm{ln} 0.23, 1)$ & $\mathcal{G}(\textrm{ln} 0.27, 1)$ &  $\mathcal{G}(\textrm{ln} 0.31, 1)$ \\
TTVs [days] & $\mathcal{G}(T_{0,c}, 0.1)$ & $\mathcal{G}(T_{0,d}, 0.1)$ & $\mathcal{G}(T_{0,b}, 0.1)$ & ---
}
\startdata
\enddata
\tablecomments{$u_1$ and $u_2$ are the limb-darkening parameters sampled following the reparameterization described by \cite{kipping13}; $P_{rot}$ is the rotation period of \sname; ln($Q_0$) is the quality factor for the secondary oscillation used to fit the stellar variability; $\Delta Q_0$ is the difference between the quality factors of the first and second modes; f is the fractional amplitude of the second mode compared the first; $\mu$ is the mean of the light curve; $\sigma_i$ and $\sigma_j$ are the light curve noise and in-transit jitter. Priors are noted for parameters that were directly sampled. The distributions are as follows -- $\mathcal{G}$: Gaussian; $\mathcal{H}$: Half-normal; $\mathcal{U}$: Uniform. $\sigma_\textrm{flux}$ is the standard deviation of the light curve.}
\end{deluxetable*}

\begin{deluxetable*}{l r r r r}[hbtp]
\tabletypesize{\footnotesize}
\tablecaption{Predicted transit midpoint events for \planete.} \label{tab:e}
\tablehead{\colhead{P [days]} & \colhead{Posterior Prob.} & \colhead{Observable Dates UT} & & \\
\hline
44.1699 $\pm$ 0.0001 & 0.071 & 21/12/2021 15:17 & 03/02/2022 19:22 & 19/03/2022 23:27 \\
\hline
45.0033 $\pm$ 0.0001 & 0.069 & 23/12/2021 07:17 & 06/02/2022 07:22 & 23/03/2022 07:27 \\
\hline
45.8687 $\pm$ 0.0001 & 0.066 & 25/12/2021 00:50 & 08/02/2022 21:41 & 26/03/2022 18:32 \\
\hline
46.7681 $\pm$ 0.0001 & 0.064 & 26/12/2021 20:00 & 11/02/2022 14:26 & 30/03/2022 08:52 \\
\hline
47.7035 $\pm$ 0.0001 & 0.062 & 28/12/2021 16:54 & 14/02/2022 09:47 & 03/04/2022 02:40 \\
\hline
48.6770 $\pm$ 0.0001 & 0.061 & 30/12/2021 15:38 & 17/02/2022 07:53 & 07/04/2022 00:07 \\
\hline
49.6911 $\pm$ 0.0001 & 0.059 & 01/01/2022 16:18 & 20/02/2022 08:54 & 11/04/2022 01:29 \\
\hline
50.7484 $\pm$ 0.0001 & 0.057 & 03/01/2022 19:03 & 23/02/2022 13:01 & 15/04/2022 06:59 \\
\hline
51.8516 $\pm$ 0.0001 & 0.055 & 06/01/2022 00:01 & 26/02/2022 20:27 & 19/04/2022 16:53 \\
\hline
53.0039 $\pm$ 0.0001 & 0.053 & 08/01/2022 07:19 & 02/03/2022 07:25 & 24/04/2022 07:30 \\
\hline
54.2085 $\pm$ 0.0001 & 0.051 & 10/01/2022 17:08 & 05/03/2022 22:09 & 29/04/2022 03:09 \\
\hline
55.4692 $\pm$ 0.0001 & 0.049 & 13/01/2022 05:39 & 09/03/2022 16:55 & --- \\
\hline
56.7899 $\pm$ 0.0001 & 0.047 & 15/01/2022 21:03 & 13/03/2022 16:00 & --- \\
\hline
58.1750 $\pm$ 0.0001 & 0.045 & 18/01/2022 15:32 & 17/03/2022 19:44 & --- \\
\hline
59.6294 $\pm$ 0.0001 & 0.042 & 21/01/2022 13:21 & 22/03/2022 04:27 & --- \\
\hline
61.1583 $\pm$ 0.0001 & 0.039 & 24/01/2022 14:44 & 26/03/2022 18:32 & --- \\
\hline
62.7678 $\pm$ 0.0001 & 0.037 & 27/01/2022 19:59 & 31/03/2022 14:25 & --- }
\startdata
\enddata
\tablecomments{Transit dates were calculated using the Transit Service Query Form on the NASA Exoplanet Archive \citep{Akeson2013}. We queried observable transits between December 3, 2021 through April 30, 2022. Dates presented in DD/MM/YYYY format. A machine-readable version of this table can be found here.}
\end{deluxetable*}

\facilities{\tess\ \citep{Ricker2015}, \textit{Kepler} \citep{Howell2014}}

\software{\texttt{exoplanet} \citep{exoplanet2021},
          \texttt{EVEREST 2.0} \citep{luger18},
          \texttt{lightkurve} \citep{lightkurve},
          \texttt{matplotlib} \citep{matplotlib},
          \texttt{PyMC3} \citep{Salvatier16},
          \texttt{starry} \citep{luger19},
          \texttt{theano} \citep{theano},
          \texttt{tica} \citep{fausnaugh20},
          \texttt{showyourwork!} \citep{luger21},
          \texttt{astropy} \citep{astropy:2013, astropy18},
          \texttt{astroquery}\citep{astroquery19},
          \texttt{numpy} \citep{numpy},
          \texttt{celerite2} \citep{dfm17},
          \texttt{batman} \citep{Kreidberg15}
          }

\bibliography{main}
\bibliographystyle{aasjournal}

\end{document}